\documentclass{aa} 
\usepackage{txfonts}
\usepackage{longtable}  
\usepackage{rotating}
\usepackage{natbib}
\usepackage{graphicx}
\usepackage{graphics}
\usepackage{psfrag}
\usepackage{amssymb}
\usepackage{xspace}
\usepackage{color}
\bibpunct{(}{)}{;}{a}{}{,}
\def\Teff{\ensuremath{T_{\mathrm{eff}}}}
\def\logg{\ensuremath{\log g}}

\def\vsini{\ensuremath{{\upsilon}\sin i}}
\def\kms{$\mathrm{km\,s}^{-1}$}

\def\espa{ESPaDOnS}

\def\M{\ensuremath{M_{\odot}}}

\def\bd{B$_{\rm d}$}

\newcommand{\bonnsai}{\mbox{\textsc{Bonnsai}}\xspace}
\begin{document} 
\title{Evidence of magnetic field decay in massive main-sequence stars} 
\subtitle{} 
\author{L. Fossati\inst{1,2}			\and
	F. R. N. Schneider\inst{3,2}		\and
	N. Castro\inst{2}			\and
	N. Langer\inst{2}			\and
	S. Sim{\'o}n-D{\'{\i}}az\inst{4,5}	\and
	A. M\"uller\inst{2}			\and
	A. de Koter\inst{6,7}			\and
	T. Morel\inst{8}			\and
	V. Petit\inst{9}			\and
	H. Sana\inst{7}				\and
	G. A. Wade\inst{10}	
}
\institute{
	Space Research Institute, Austrian Academy of Sciences, Schmiedlstrasse 		6, A-8042 Graz, Austria\\
	\email{luca.fossati@oeaw.ac.at}
	\and
	Argelander-Institut f\"ur Astronomie der Universit\"at Bonn, Auf dem 			H\"ugel 71, 53121, Bonn, Germany 
	\and
	Department of Physics, Denys Wilkinson Building, Keble Road, Oxford, OX1 	3RH, United Kingdom
	\and
	Instituto de Astrof\'isica de Canarias, 38200, La Laguna, Tenerife, 			Spain
	\and
	Departamento de Astrof\'isica, Universidad de La Laguna, E38205, La 			Laguna, Tenerife, Spain
	\and
	Astronomical Institute Anton Pannekoek, Amsterdam University, Science 			Park 904, 1098 XH, Amsterdam, The Netherlands
	\and
	Instituut voor Sterrenkunde, Universiteit Leuven, Celestijnenlaan 200 D, 	B-3001 Leuven, Belgium
	\and
	Institut d'Astrophysique et de G\'eophysique, Universit\'e de Li\`ege, 			All\'ee du 6 Ao\^ut, B\^at. B5c, 4000 Li\`ege, Belgium
	\and
	Department of Physics and Space Sciences, Florida Institute of 				Technology, Melbourne, FL 32904, USA
	\and
	Department of Physics, Royal Military College of Canada, PO Box 17000 			Station Forces, Kingston, ON K7K 7B4, Canada
}
\date{} 
\abstract
{A significant fraction of massive main-sequence stars show strong, large-scale magnetic fields. The origin of these fields, their lifetimes, and their role in shaping the characteristics and evolution of massive stars are currently not well understood. We compile a catalogue of 389 massive main-sequence stars, 61 of which are magnetic, and derive their fundamental parameters and ages. The two samples contain stars brighter than magnitude 9 in the $V$ band and range in mass between 5 and 100\,\M. We find that the fractional main-sequence age distribution of all considered stars follows what is expected for a magnitude limited sample, while that of magnetic stars shows a clear decrease towards the end of the main sequence. This dearth of old magnetic stars is independent of the choice of adopted stellar evolution tracks, and appears to become more prominent when considering only the most massive stars. We show that the decreasing trend in the distribution is significantly stronger than expected from magnetic flux conservation. We also find that binary rejuvenation and magnetic suppression of core convection are unlikely to be responsible for the observed lack of older magnetic massive stars, and conclude that its most probable cause is the decay of the magnetic field, over a time span longer than the stellar lifetime for the lowest considered masses, and shorter for the highest masses. We then investigate the spin-down ages of the slowly rotating magnetic massive stars and find them to exceed the stellar ages by far in many cases. The high fraction of very slowly rotating magnetic stars thus provides an independent argument for a decay of the magnetic fields.}
\titlerunning{Evidence for magnetic field decay in massive stars}
\authorrunning{L. Fossati et al.}
\maketitle
\section{Introduction}\label{introduction}
Magnetic fields are an important constituent of astrophysical plasmas. They play a vital role in all types of stars \citep{chanmugam1992,donati2009} and may affect their internal structure and evolution \citep[e.g.,][]{jackson2009,meynet2011,langer2012,petermann2015}. Whereas only a few magnetic massive stars were known until recently, such as HD\,37022 \citep{donati2002}, HD\,191612 \citep{donati2006}, $\tau$\,Sco \citep{donati2006b}, and  HD\,37742 \citep{bouret2008}, new intense searches have revealed many more magnetic O- and B-type stars \citep[e.g.,][]{grunhut2009,petit2013,hubrig2014,fossati2015,fossati2015b,castro2015}, implying an incidence fraction of strong, large-scale magnetic fields in massive stars of about ten percent \citep{grunhut2012,wade2015a}.

While the origin of these potentially long-lived magnetic fields in massive stars is still under debate, they are thought to be fossil in the sense that the stars do not host the required conditions for a dynamo process that could generate fields in situ \citep{donati2009}, that is, for convection and rapid rotation \citep[Donati \& Landstreet\,2009; Cantiello et al.\,2009; but see][]{potter2012}. The strongest evidence for a fossil origin of magnetic fields in intermediate-mass stars is the lack of correlations of the magnetic field strength with rotation period (from 0.5\,d to about 100\,yr) and mass \citep[from 1.5 to about 5\,\M; e.g.,][]{oleg2006}. Recent studies show the same for higher mass stars \citep[][]{wade2015a}. Their envelopes are generally radiative, with convection occurring only in a very low-mass subsurface region \citep{cantiello2009,sanyal2015}. The magnetic stars among them are often very slow rotators, although a few magnetic rapid rotators have been found \citep{petit2013}. 

\citet{braithwaite2004} found that in main-sequence (MS) A-type stars magnetic field configurations, which later on have been found to resemble those observed in magnetic O- and B-type stars, can indeed be stable, in the sense that their decay may take longer than the stellar lifetime. For such intermediate-mass stars, several studies have attempted to characterise the magnetic field properties as a function of mass and age \citep[e.g.,][]{hubrig2000,oleg2006,landstreet2007,landstreet2008}. For example, \citet{hubrig2000,hubrig2007} and \citet{oleg2006} proposed that field chemically peculiar (CP) magnetic A-type (Ap) stars concentrate primarily in the middle of their MS evolution, particularly for stars with masses $\leq$2\,\M. On the basis of a large sample of open cluster CP stars, \citet{landstreet2007,landstreet2008} showed that the surface magnetic field of CP stars more massive than 3\,\M\ may decline with age at a rate higher than expected by flux conservation.

A comparison of the position of magnetic O- and B-type stars in the Hertzsprung-Russell diagram (HRD) with evolutionary tracks \citep[see e.g., Fig.~2 of][]{fossati2015} indicates that the vast majority of these stars are still on the MS, meaning that they are undergoing core hydrogen burning. The only potential exception is the blue supergiant HD\,37742 \citep[an Orion belt star that has a very weak magnetic field][]{bouret2008,blazere2015}, which is also identified as a core hydrogen burning star by the evolutionary tracks of \citet{brott2011}. The reasons for this dearth of evolved magnetic stars is currently not understood, although the expansion of the radius and the speed of the evolution in the blue supergiant phase could play a role, making magnetic field detections more challenging. 

To investigate the evolution of magnetic fields in massive stars in more detail, we have derived the probability density function (PDF) of fractional MS ages, $\tau$, of the magnetic O- and B-type stars (which we call $\tau$-distribution for brevity and clarity from here on), and compared it to that of a large comparison sample of Galactic massive stars. In Sect.~\ref{sec:sample}, we present the investigated stellar samples and the derivation of the stellar parameters. In Sect.~\ref{sec:results} we investigate the $\tau$-distribution of magnetic and non-magnetic O- and B-type stars, and find significant differences. We discuss possible origins of these differences in Sect.~\ref{sec:discussion} and conclude in Sect.~\ref{sec:conclusion}.
\section{Stellar sample and analysis}\label{sec:sample}
Our sample of magnetic massive stars comprises the compilation of \citet{petit2013}, with the addition of magnetic stars presented in later publications: HD\,133518, HD\,147932 \citep{alecian2014}, HD\,47887 \citep{fossati2014}, HD\,44743, HD\,52089 \citep{fossati2015}, HD\,54879 \citep{castro2015}, HD\,43317 \citep{briquet2013}, HD\,1976 \citep{neiner2014}, CPD\,$-$57\,3509 \citep{norbert2015}, HD\,23478, HD\,345439 \citep{sikora2015,hubrig2015}, and CPD\,$-$62\,2124 (Castro et al. in prep.). Because of the lack of atmospheric stellar parameters, we did not consider the magnetic binaries reported by \citet{alecian2014} and \citet{hubrig2014}. In total, the sample consists of 76 magnetic stars. For each star, we considered $V$-band magnitude, dipolar magnetic field strength (\bd), effective temperature (\Teff), surface gravity (\logg), and projected rotational velocity (\vsini). When possible, we also further updated the stellar properties. Specifically for HD\,47777 we used properties from \citet{fossati2014} and for CPD\,$-$28\,2561 from \citet{wade2015b}.

Our comparison sample is the same as used by \citet{castro2014} and references therein, which is mostly based on the IACOB\footnote{{\tt www.iac.es/proyecto/iacob}} spectroscopic survey of Northern Galactic O- and B-type stars \citep{sergio2011a,sergio2011b,sergio2015}. This comparison sample, which also contains the magnetic stars, consists of 560 stars for which we assume here that it is dominated by non-magnetic stars even though many of these stars have not yet been investigated for magnetic fields. For each star we considered the $V$-band magnitude from SIMBAD and the \Teff\ and \logg\ values used by \citet{castro2014}. Since this sample also contains the magnetic stars, we avoided duplication in these cases and used the atmospheric parameters given in the papers dedicated to the magnetic field detection or characterisation. A thorough discussion of the biases that may be present in this sample is given by \citet{castro2014} and we return to this point in Sect.~\ref{sec:results}.

To have a magnitude-limited sample of stars that is as complete and representative as possible, we decided to restrict the analysis to stars brighter than $V$\,=\,9\,mag. The benefits of such a sample are twofold. On the one hand, it allows us to statistically compare the sample of magnetic stars to that of all stars and, on the other hand, it facilitates comparisons with population synthesis models. We note that we did not consider stellar luminosities because reliable parallaxes are currently not available for stars fainter than $V$\,$\approx$\,7\,mag \citep{vanleeuwen2007}.

We used the Bayesian tool \bonnsai\footnote{The \bonnsai\ web-service is available at {\tt http://www.astro.uni-bonn.de/stars/bonnsai}.} to determine the full posterior probability distribution of stellar mass (both initial and present-day), radius, luminosity, age, and fractional MS age for each star \citep[see][for more details]{fabian2014}. To that end, we simultaneously matched the \Teff\ and \logg\ values for each star to the rotating Milky Way stellar evolution models of \citet{brott2011}. We used the Salpeter initial mass function \citep{salpeter1955} as initial mass prior and a Gaussian initial rotational velocity prior centred on 100\,\kms\ with a half-width at half-maximum of 125\,\kms\ \citep{hunter2008}.

\begin{figure}[]
\includegraphics[width=85mm,clip]{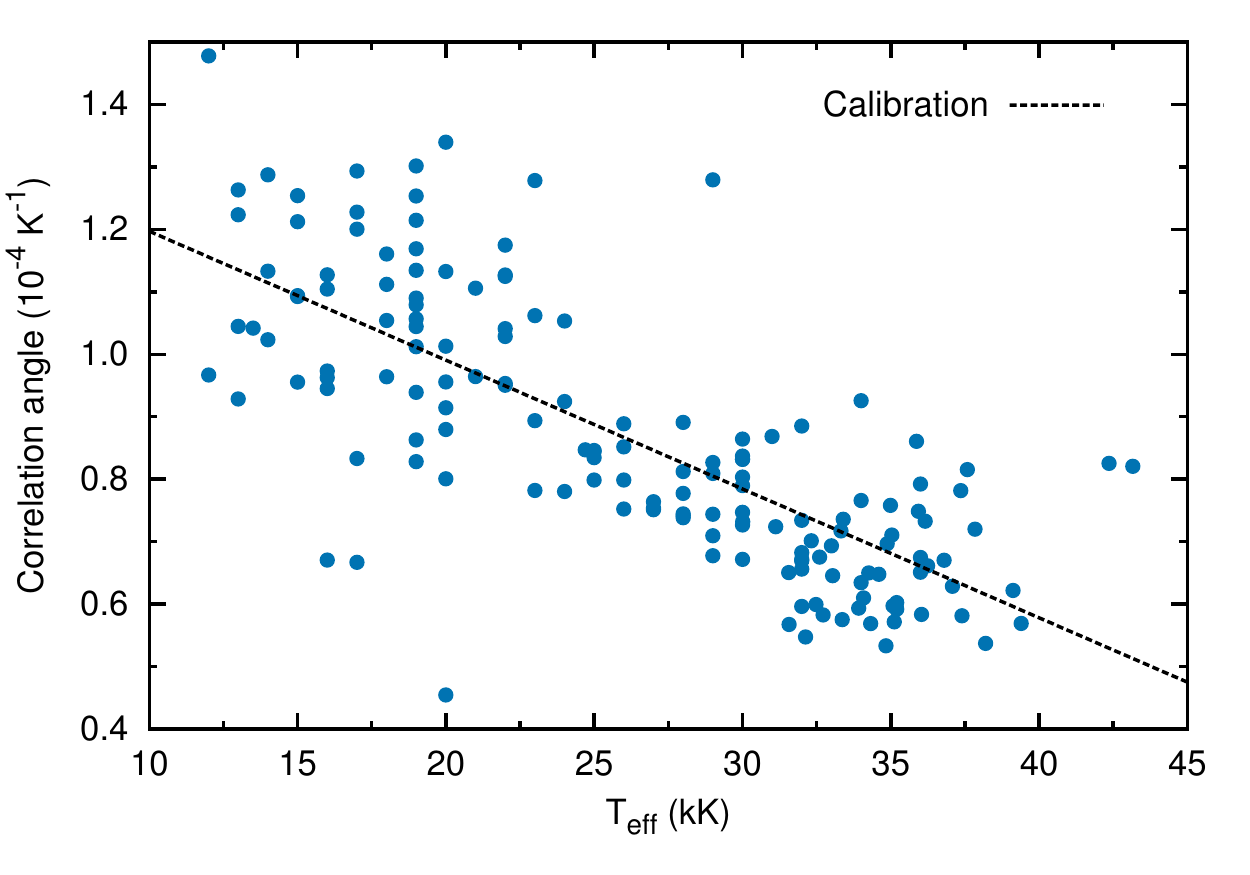}
\caption{Correlation parameters inferred for 157 stars from our sample and the resulting linear calibration (Eq.~\ref{eq:calibration-correlation-parameter}).}
\label{fig:calibration-correlation-parameter}
\end{figure}

In nearly all considered sources, the stellar atmospheric parameters (\Teff\ and \logg) were determined by fitting observed spectra. The atmospheric parameters are therefore correlated in the sense that a hotter or cooler temperature requires a higher or lower gravity to obtain a similarly good fit to the spectrum. Neglecting such correlations may lead to severely over- or under-estimated mass and age uncertainties (by up to factors of 2) and significantly biased parameters \citep[by up to 50--80\% of the $1\sigma$ error bars;][]{fabian2015}. To avoid this, we incorporated correlations between \Teff\ and \logg\ in \bonnsai\ through the covariance matrix as formulated by \citet{fabian2015}. These authors provided a parametrisation of the covariance matrix that relies on conventional, that is, marginalised, $1\sigma$ uncertainties for \Teff\ and \logg\ and a correlation parameter, $\varphi$, that describes how \Teff\ and \logg\ co-vary. This parametrisation only depends on the method with which the observables have been determined. The advantage of this parametrisation of the covariance matrix is that correlations can be accounted for even if the exact correlations are not known in each individual case. We determined the correlation parameter for 157 stars in our sample by fitting tilted Gaussian functions to $\chi^2$-maps in the \Teff\ and \logg\ plane (Fig.~\ref{fig:calibration-correlation-parameter}). For the remaining stars, the parameter space is either sampled too coarsely to allow for a reliable determination of the correlation parameter with the methods and atmosphere grids described by \citet[][see also Lefeveret al.\ 2010]{castro2012} or the correlations are not published together with the stellar parameters (this is for example the case for the O-type stars in the IACOB sample for which the publication of the complete results is still pending). The sample of stars for which we can derive the correlation parameter does however cover the whole range of \Teff\ and \logg\ values of our overall sample. We obtain (the straight line in Fig.~\ref{fig:calibration-correlation-parameter})
\begin{equation}
\varphi(\Teff) \equiv \frac{{\rm d}\logg}{{\rm d}\Teff} = \left(1.4030 \times 10^{-4} - 2.0618 \times 10^{-9} \cdot \frac{T_\mathrm{eff}}{\mathrm{K}}\right)\,\mathrm{K}^{-1}.
\label{eq:calibration-correlation-parameter}
\end{equation}
The correlation parameter $\varphi$ can be understood in the following way: when the best-fitting atmosphere model is found, that is, the best-fitting \Teff\ and \logg\ amongst other parameters, the correlation parameter $\varphi$ describes by how much the surface gravity (${\rm d}\logg$) needs to be adjusted to re-obtain the best-fitting spectrum for an effective temperature deviating by ${\rm d}\Teff$ from the global best fit. We applied this calibration to all stars in our sample by considering that the error ellipse in the \Teff--\logg\ plane is titled by the quantity given in Eq.~\ref{eq:calibration-correlation-parameter} and ensuring that the conventional error bars stay constant. We note that this calibration of the correlation parameter is only valid for analyses as those done by \citet{castro2012} because the details of the spectral analysis (the applied atmosphere code, lines used in the analysis, weights set for fitting different lines, number of free parameters, etc.) influence the exact correlations \citep[see][for more details]{fabian2015}.

\begin{figure}[]
\vspace{-3cm}
\includegraphics[width=85mm,clip]{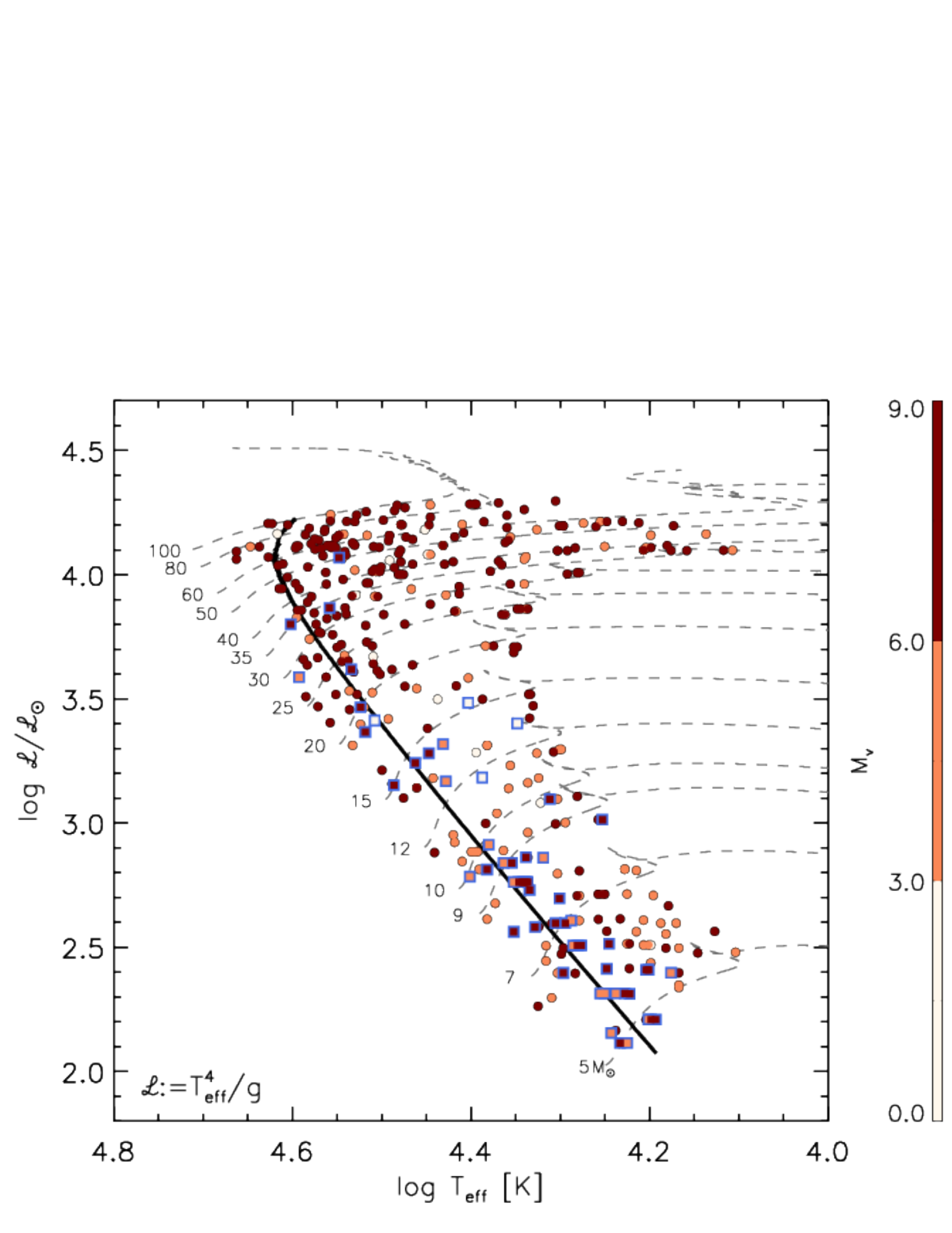}
\caption{Spectroscopic HRD of the stars considered in this work and evolutionary tracks for non-rotating stars (blue dashed lines) by \citet{brott2011}. The solid line shows the position where $\tau$\,=\,0.5. The magnetic stars are marked by blue squares. The stars are colour-coded according to their $V$-band magnitude.}
\label{fig:shrd}
\end{figure}

From the considered samples, we removed the stars that have evolved beyond the MS in the \citet{brott2011} stellar evolution tracks ($\tau$ is not defined for post-MS stars). Only one star, HD\,48099, falls on the hot side of the zero-age main sequence, and we therefore removed it from the sample. We furthermore excluded stars for which the determined initial mass is $<$5\,\M, and is within $1\sigma$ of the upper stellar model grid boundary of 100\,\M. In the end, the full sample is composed of 389 stars, with 61 objects belonging to the sample of magnetic stars. Figure~\ref{fig:shrd} shows the distribution of the two samples of stars in the spectroscopic HRD \citep{langer2014}.
%
\section{Results}\label{sec:results}
To derive the $\tau$-distribution of our samples, we summed the $\tau$-distributions of the individual stars and re-normalised the resulting PDF by the number of stars in the sample. This approach ensures that observational uncertainties of individual stars are properly taken into account. A second source of uncertainty is introduced by the finite size of the stellar samples (stochastic sampling) because our further analysis relies on the fact that the two stellar samples are representative of the true, underlying stellar populations. To quantify the significance of the $\tau$-distributions with respect to the sample size $N$, we computed bootstrapped $1\sigma$ estimates as the standard deviations of 10\,000 realisations of the $N$ combined fractional MS $\tau$-distributions where stars have been randomly sampled with replacement.
 
\begin{figure}[]
\includegraphics[width=85mm,clip]{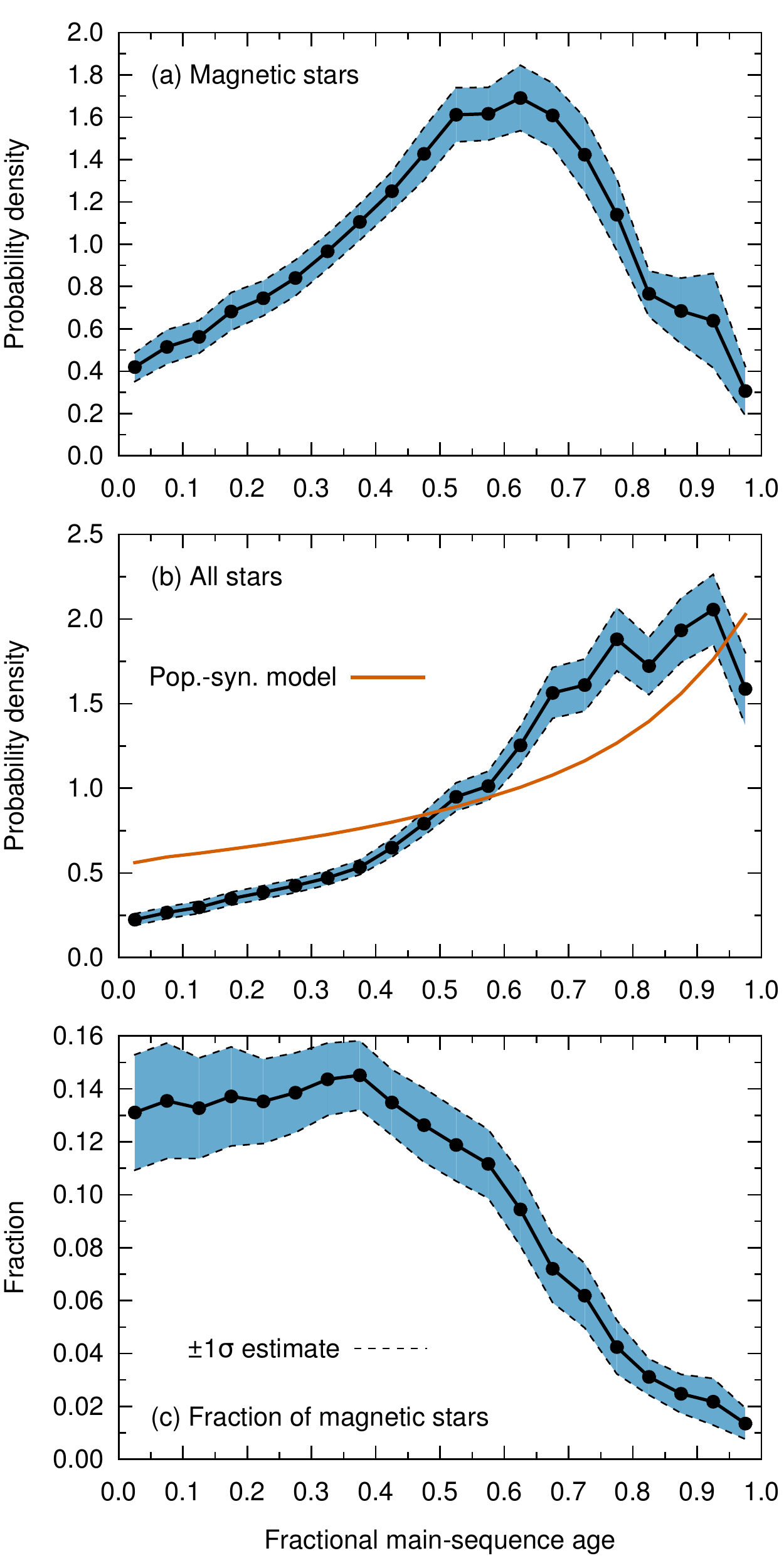}
\caption{Fractional MS $\tau$-distributions of (a) the magnetic stars and (b) all stars in our sample. Panel (c) shows the fraction of magnetic stars as a function of fractional MS age, normalised such that the incidence of magnetic stars is 7\% \citep{wade2014,fossati2015b}. The shaded regions indicate bootstrapped 1$\sigma$ estimates to give an indication of the statistical significance of the variability in the $\tau$-distributions. The solid line in the middle panel shows the $\tau$-distribution of our synthetic population of a magnitude-limited sample of massive stars (see text).}
\label{fig:age-distribution}
\end{figure}

The top and middle panel of Fig.~\ref{fig:age-distribution} show the $\tau$-distribution of the magnetic stars and that of all stars (magnetic and non-magnetic), respectively. The $\tau$-distribution of the magnetic stars increases initially, reaches a maximum at $\tau$\,$\approx$\,0.6, and drops thereafter. Hence, the bulk of magnetic stars in our sample have most probable fractional main sequence ages corresponding to the middle of their MS evolution, in agreement with what is observed for intermediate-mass CP stars \citep{hubrig2000,oleg2006,hubrig2007}.

Although this sample contains all magnetic massive stars known to date, there may be some biases and selection effects. There are two types of biases that should be considered. First, we already know from specific indicators that certain stars host magnetic fields (e.g. Of?p and He-strong stars) and there is the tendency to search preferentially for magnetic fields in stars displaying these indicators. A set of secondary indicators is present as well, such as slow rotation, peculiar emission lines in the spectra, peculiar chemistry, and hard X-ray emission, but they are not necessarily accompanied by any evidence of a magnetic field \citep[see, e.g.][in the case of the nitrogen excess]{wade2015c}, such that their contribution to this particular bias is probably small. Second, there may be a bias introduced by systematic differences in sensitivity (e.g. faint versus bright stars, slow versus rapid rotators, hot versus cool stars). However, within the MiMeS sample, the sensitivity obtained for example for slow and fast rotators is comparable \citep{wade2016}. Nevertheless, we tried to account for these biases in the interpretation of the results (Sect.~\ref{sec:discussion}). 

The main type of bias that would strongly affect the $\tau$-distribution, shown in the top panel of Fig.~\ref{fig:age-distribution}, is time dependent: there might be preferential ages associated with different samples of preferentially detected stars. For example, there may be some age dependence of the Of?p phenomenon as the young O-type star $\theta^1$\,OriC does not show Of?p signatures (but this will remain speculative until a much larger sample of magnetic O-type stars has been studied). The reality is that we do not know of any age-dependent bias present in the sample. It is important to also consider that our sample of magnetic stars is predominantly based on the compilation of \citet{petit2013}, which is mostly composed of stars detected to be magnetic with different instruments and techniques, such that selection effects, if present, are probably very complex. Since we considered all known magnetic stars, this sample is as complete as it can possibly be to the best of our current knowledge.

The $\tau$-distribution of all stars increases with fractional MS age and reveals that our sample is dominated by relatively old stars. About 70\% of the stars are in the second half of their MS evolution. The drop in $\tau$-distribution around $\tau$\,=\,1.0 arises because we did not consider any star beyond the terminal-age main sequence (TAMS) as predicted by \citet{brott2011}. This means that some stars that overlap the MS within their error bars are missing in the $\tau$-distribution because their best-fitting age is beyond the TAMS. 

To understand the characteristics of the $\tau$-distribution of all stars, we computed a synthetic population of massive stars, assuming continuous star formation, the Salpeter initial mass function \citep{salpeter1955}, and a uniform distribution of heliocentric distances. We then set the magnitude cut at $V$\,=\,9\,mag and derived the $\tau$-distribution of the synthetic stars, which shows a behaviour similar to the observed one (Fig.~\ref{fig:age-distribution}b). The increase in $\tau$-distribution with increasing fractional MS age is a bias caused by the fact that stars become more luminous as they age: a magnitude-limited sample of MS stars contains a larger portion of old stars than a similar volume-limited sample \citep[cf. Malmquist bias;][]{malmquist}. Because of the same magnitude cut, the $\tau$-distribution of magnetic stars is also affected in the same way by this bias.

Some differences between the observed and synthetic $\tau$-distributions of all stars are present. These are probably caused by biases in the sample selection. For instance, the IACOB sample of B-type stars is slightly biased towards older stars for the study of macroturbulence, while for the O-type stars in the northern hemisphere it is complete up to $V$\,=\,9\,mag. Simplifying assumptions in the population synthesis model (e.g., uniform distance distribution across the whole sky) may also contribute. In addition, Fig.~\ref{fig:shrd} shows that for masses higher than 30\,\M\ the region close to the ZAMS is not sampled by the observations, as previously noted by \citet{castro2014}. The lack of very massive stars close to the ZAMS slightly affects the $\tau$-distribution of all stars and contributes to the discrepancy with the synthetic $\tau$-distribution at $\tau$\,$<$\,0.5. Part of the discrepancy at older ages ($\tau$\,$>$\,0.5) may arise because the models do not consider a mass-dependent overshooting, while observations suggest that it may be present \citep{castro2014}.

The bottom panel of Fig.~\ref{fig:age-distribution} shows the fraction of magnetic stars as a function of fractional MS age. Our current samples are not complete, that is they do not contain all (magnetic) massive stars brighter than $V$\,=\,9\,mag. In addition, our sample is biased towards magnetic stars, as indicated by the fact that the number ratio of magnetic and all stars is 16\%, which is higher than what is found by dedicated surveys \citep[7\%;][]{wade2014,fossati2015b}. Therefore, to also facilitate direct comparisons with future observations, we re-scaled the fraction of magnetic massive stars in Fig.~\ref{fig:age-distribution}c such that the overall magnetic incidence is 7\%\footnote{Let $\left(\mathrm{d}p/\mathrm{d}\tau\right)_\mathrm{mag}$ and $\left(\mathrm{d}p/\mathrm{d}\tau\right)_\mathrm{all}$ be the $\tau$-distributions of magnetic and all stars, respectively. To compute the incidence of magnetic stars as a function of fractional MS age, $f_\mathrm{mag}(\tau)$, we need to compute the ratio of the numbers of magnetic and all stars for each $\tau$-bin: $f_\mathrm{mag}(\tau) = N_\mathrm{mag} \left(\mathrm{d}p/\mathrm{d}\tau\right)_\mathrm{mag} / N_\mathrm{all} \left(\mathrm{d}p/\mathrm{d}\tau\right)_\mathrm{all} = N_\mathrm{mag}/N_\mathrm{all} \left(\mathrm{d}p/\mathrm{d}\tau\right)_\mathrm{mag} / \left(\mathrm{d}p/\mathrm{d}\tau\right)_\mathrm{all}$ where $N_\mathrm{mag}$ and $N_\mathrm{all}$ are the total numbers of magnetic and all stars, respectively, such that $N_\mathrm{mag}/N_\mathrm{all}\approx$7\% is the overall incidence of magnetic massive stars.} \citep{wade2014,fossati2015b}. Nevertheless, this is not a problem as long as the samples are representative, hence leading to $\tau$-distributions comparable to those obtained from complete samples. Assessing how representative a sample is for the complete population can only be done statistically because the complete population will never be known. In our case, the robustness of the $\tau$-distributions is given by bootstrapped $1\sigma$-estimates. These bootstrapped uncertainties quantify the variations in our $\tau$-distributions because of stochastic sampling from a parent distribution and will further decrease when larger samples become available in the future. The similar trend in $\tau$-distributions of our sample stars and the population synthesis model (which is by definition complete) also indicates that our sample of stars can be considered representative of all stars brighter than $V$\,=\,9\,mag.

For stars younger than $\tau$\,$\approx$\,0.5$-$0.6, the fraction of magnetic massive stars is, within the error bars, consistent with being constant. For older stars, the fraction of magnetic massive stars decreases steeply with increasing $\tau$. Towards the TAMS only about 2\% of massive stars are detected to have large-scale magnetic fields, whereas about 14\% show such fields towards the ZAMS. Hence, the fraction of magnetic stars drops by about a factor of seven between the ZAMS and the TAMS\footnote{Because it is based on the ratio of two PDFs, this result does not depend on the star formation history, assuming that it is the same for the magnetic and non-magnetic stars.}.

\begin{figure}[]
\includegraphics[width=85mm,clip]{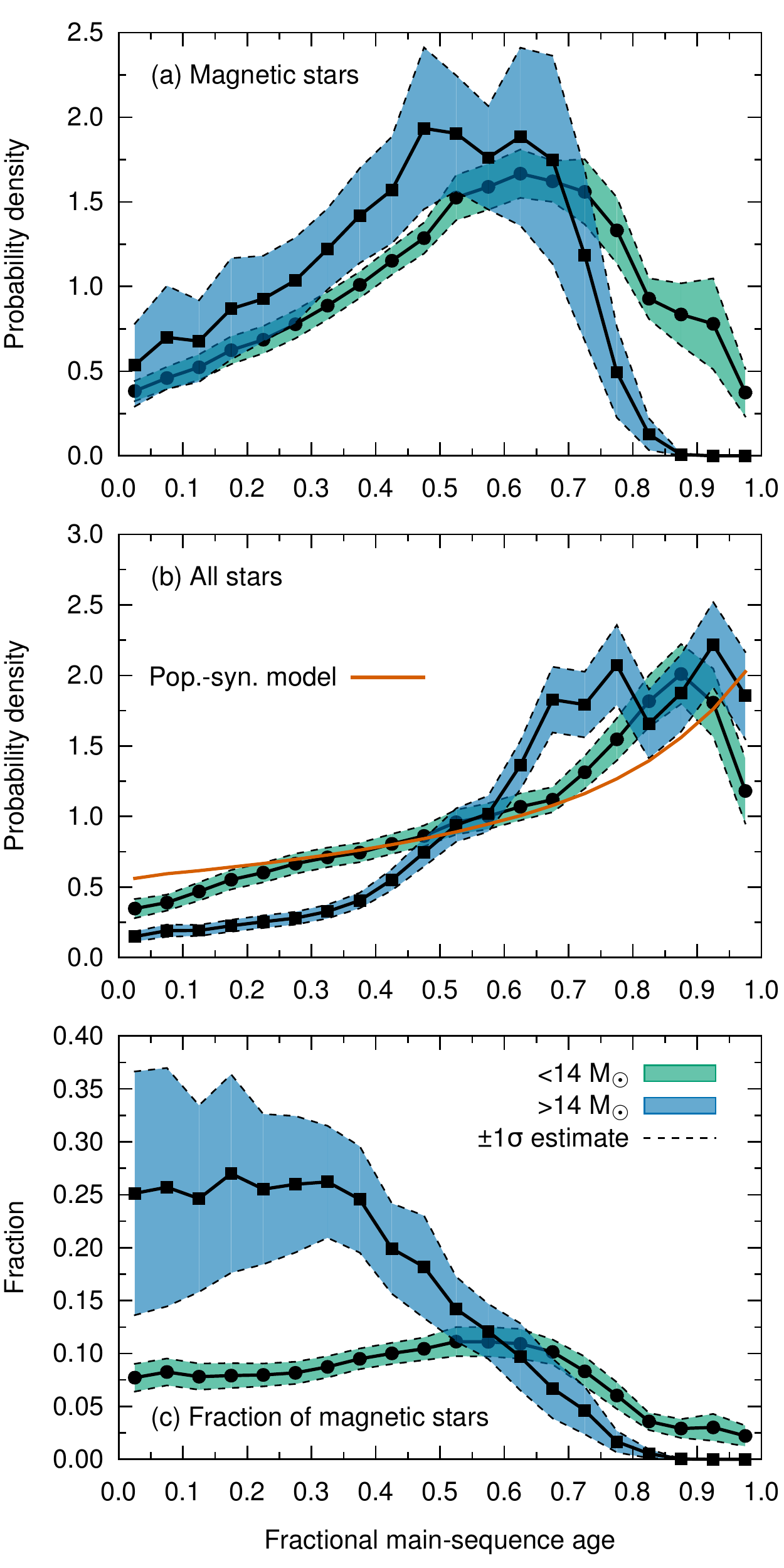}
\caption{Same as Fig.~\ref{fig:age-distribution}, but for subsets of stars less and more massive than 14\,\M.}
\label{fig:age-distribution-2}
\end{figure}

To investigate whether the dearth of old magnetic stars depends on mass, we divided our stellar sample into subsamples of stars less and more massive than 14\,\M\ and computed the $\tau$-distributions for both samples (Fig.~\ref{fig:age-distribution-2}). The sample is split such that the high-mass sample is as massive as possible to easily spot a potential mass dependence, while still containing sufficient stars to obtain a reliable $\tau$-distribution. The chosen mass cut at 14\,\M\ leads to 12 magnetic stars in the high-mass sample and 50 in the low-mass sample (in total there are 145 and 246 stars, respectively; the best-fitting initial mass of one star is exactly 14\,\M\ and we included this star in both samples). The ratios of the $\tau$-distributions (bottom panel of Fig.~\ref{fig:age-distribution-2}) were re-normalised to an incidence rate of 7\%, as in the bottom panel of Fig.~\ref{fig:age-distribution}. The difference between the two curves at young ages is due to the lack of young massive stars in our sample, as discussed above. Because of these issues, the incidence rates at young ages in the high-mass sample in Fig.~\ref{fig:age-distribution-2} are most likely affected by biases and should not be taken at face value.

We find that the $\tau$-distributions of the low- and high-mass samples show a similar increasing trend to that of the combined sample and also to that of our basic population synthesis model (see middle panel of Fig.~\ref{fig:age-distribution-2}). The differences between the observed and synthetic $\tau$-distributions of all stars (middle panel of Fig.~\ref{fig:age-distribution-2}) close to the ZAMS reflect the lack of young massive stars of our sample, as pointed out earlier in this section and as noted by \citet{castro2014}. However, the $\tau$-distributions of the magnetic stars show differences at roughly the 2$\sigma$ level close to the TAMS (see top panel of Fig.~\ref{fig:age-distribution-2}). There the $\tau$-distribution of the higher mass sample drops more steeply than that of the lower mass sample, and this drop may start at younger fractional MS age. This indicates that the relative deficiency of evolved magnetic stars may be mass dependent, or in other words, more pronounced in more massive stars.

The stellar models of \citet{brott2011} used in this study employ a larger convective core overshooting than the models of \citet{ekstrom2012}. A larger overshooting results in systematically younger ages because the TAMS is pushed to cooler temperatures. A comparison of the \citet{brott2011} and \citet{ekstrom2012} models in the HRD \citep[e.g.,][]{castro2014} reveals that there is a nearly constant offset between points of equal fractional MS ages for initial masses in the range 10--40\,\M, and we would have derived $\tau$ values lower by about 0.1 if we had used the \citet{ekstrom2012} models. A restriction of the two samples of stars to the 10--40\,\M\ range shows that such an offset does not significantly influence the shape of the $\tau$-distributions shown here and therefore our findings.
\section{Discussion}\label{sec:discussion}
In the previous section we established that there is a dearth of relatively old magnetic MS massive stars. A similar trend was found by \citet{landstreet2007,landstreet2008} on the basis of a sample of magnetic A- and B-type stars in open clusters. \citet{landstreet2007,landstreet2008} suggested that for stars more massive than 3\,\M\ the magnetic field strength may decrease at a rate faster than that predicted by flux conservation. We have identified four scenarios that may explain the deficit of evolved magnetic massive stars: {\it i}) magnetic flux conservation, {\it ii}) binary rejuvenation, {\it iii}) magnetic suppression of core convection, and {\it iv}) magnetic field decay.
\subsection{Magnetic flux conservation}\label{sec:flux_conservation}
While they are on the MS, stars expand with time, resulting in progressively weaker surface magnetic fields for a conserved magnetic flux, $\phi$\,=\,B$_{\rm d}$R$^2$. As a consequence, the surface magnetic field strength for old stars may fall below the observational detection threshold and cause an apparent dearth of evolved magnetic stars.

In Sect.~\ref{sec:results} we concluded that  only one in seven young magnetic massive stars would still be detectable as magnetic after it had reached the TAMS. To determine whether this is caused by the decrease in the surface magnetic field expected as a result of flux conservation, we estimated the fraction of ZAMS magnetic massive stars that would still be detectable as magnetic after they had reached the TAMS, assuming the surface magnetic field strength evolves only as a result of flux conservation.

\begin{figure}[]
\vspace{-4.5cm}
\includegraphics[width=85mm,clip]{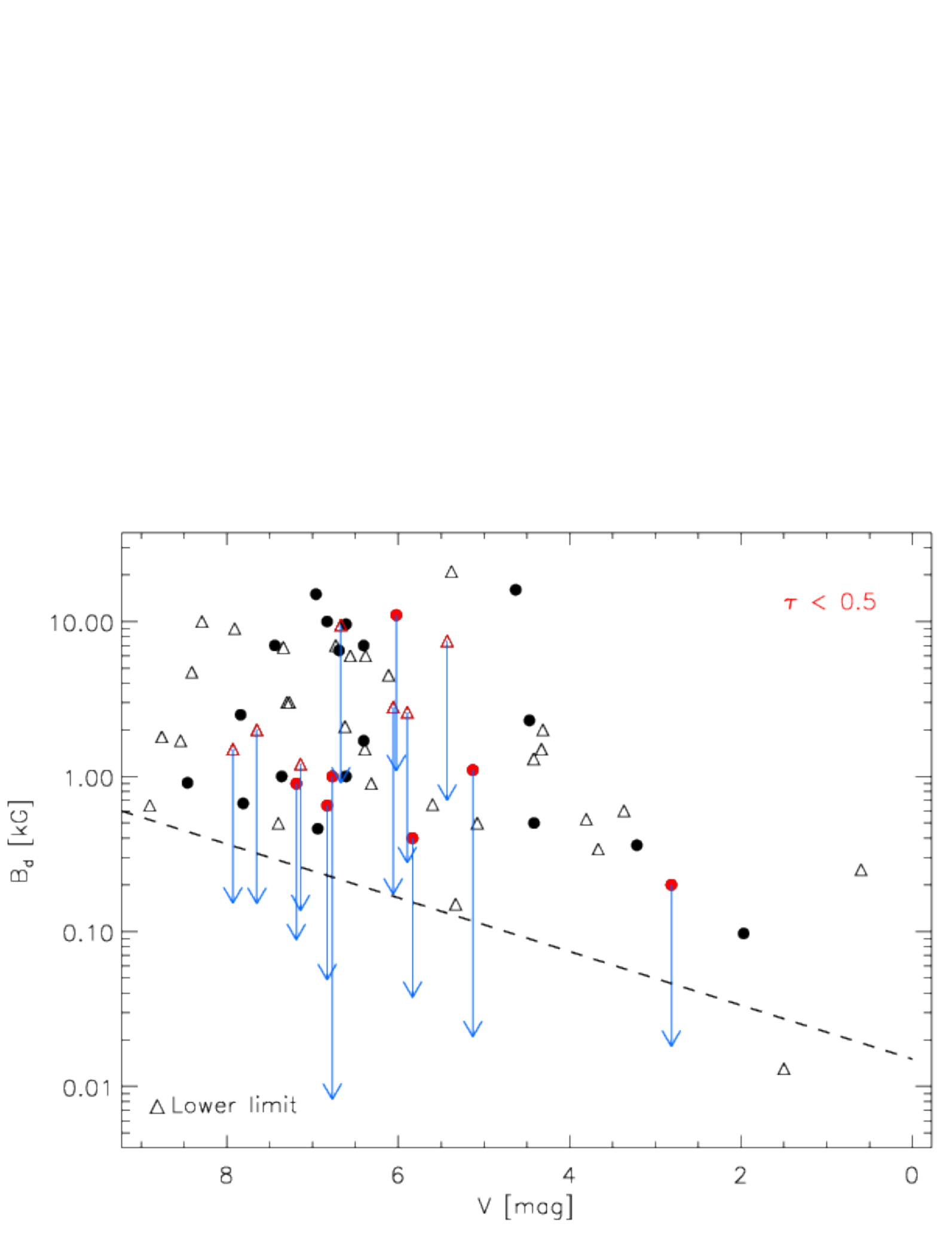}
\caption{Dipolar magnetic field strength as a function of apparent $V$-band magnitude for the magnetic massive stars. The dashed line indicates our estimated empirical detection threshold. Open triangles indicate lower limits on \bd; red symbols show the position of the stars having a fractional MS age smaller than 0.5. Arrows point to the \bd\ values that young stars (i.e., $\tau<$0.5) would have on the TAMS assuming flux conservation.}
\label{fig:evolution-detection}
\end{figure}

Figure~\ref{fig:evolution-detection} shows the measured dipolar magnetic field strengths (\bd) of the magnitude-limited sample of magnetic massive stars as a function of their $V$-band magnitude. From this plot we derive a rough indication of the current capability of magnetic field detections in massive stars as a function of apparent magnitude. Following the lower boundary of the bulk of points, we outlined an empirical detection threshold with the dashed line.

Although the detection of a stellar magnetic field depends on a number of factors, such as the adopted instrument (high- or low-resolution spectropolarimeter), observed wavelength range, detection technique, and characteristics of the star (e.g., number of available spectral lines and their width\footnote{The majority of the magnetic massive stars presents rather narrow lines \citep[see e.g., Table~1 of][]{petit2013}.}), Fig.~\ref{fig:evolution-detection} indicates that the stellar magnitude plays an important role. This is probably because most of the confirmed magnetic massive stars have so far been detected, and characterised, using similar instruments or telescopes (e.g. \espa\ and HARPSpol both on 4\,m class telescopes) and analysis methods \citep[i.e. Least-Squares Deconvolution;][]{donati1997,kochukhov2010}. 

In Fig.~\ref{fig:evolution-detection} we highlight the stars with the most probable fractional MS age $<$\,0.5 and consider these as the ``young'' stars sample. To determine the expansion factor of the radius of each star, we considered the ratio between the radius at the TAMS, as given by the evolutionary tracks of \citet{brott2011}, and the radius of the young stars at the best-fitting age minus 1$\sigma$, to be conservative (on average the ratio between the two radii is $\approx$3). We then calculated the dipolar magnetic field strength that each young star would have on the TAMS assuming the field strength decreases only because of flux conservation. This value is indicated in Fig.~\ref{fig:evolution-detection} by the endpoint of each arrow.

The sample of magnetic stars is composed of 14 young stars\footnote{We note that a very high percentage of stars has a best-fitting $\tau$ just above 0.5 as shown in Fig.~\ref{fig:shrd}.}. By following the evolution of the magnetic field considering flux conservation, at least 5 stars would still be detectable as magnetic on the TAMS. This is about one-third of the sample, to be compared to a fraction of one-seventh derived from Fig.~\ref{fig:age-distribution}. This indicates that flux conservation alone is probably not the origin of the rapid decrease of the fraction of magnetic stars at $\tau$\,$\gtrsim$\,0.6. We performed various tests for the reliability of this result and found that the fraction of stars still detectable as magnetic at the TAMS is independent of the magnitude cut and the fractional MS age adopted to identify the young stars (i.e. $\tau$\,$<$\,0.4 or 0.6). 

Our result is strengthened by the fact that the factor of three derived from Fig.~\ref{fig:evolution-detection} is an upper limit because {\it i}) the empirical detection threshold drawn in Fig.~\ref{fig:evolution-detection} is an upper limit, as it is indeed possible to detect magnetic fields smaller than what is indicated in the plot \citep[e.g.,][]{fossati2015}; {\it ii}) the \bd\ value available for some of the young stars is a lower limit; {\it iii}) we adopted a larger radius expansion factor\footnote{Here, the radius expansion factor is the ratio between the stellar radii at the TAMS and at the best-fitting age.} compared to the best fit; {\it iv}) stars become brighter as they age, hence increasing the chances to detect them as magnetic closer to the TAMS; {\it v}) stars slow and cool down along the MS making magnetic fields easier to detect close to the TAMS; {\it vi}) the use of a different set of stellar evolution tracks, for instance that of \citet{ekstrom2012}, would lead to lower radius expansion factors \citep[i.e., shorter arrows in Fig.~\ref{fig:evolution-detection}; ][]{sanyal2015}. Each of these considerations would lead to an increase in the number of young stars that would still be detectable as magnetic at the TAMS, assuming flux conservation.
\subsection{Binary rejuvenation}\label{sec:binary_rejuvenation}
The origin of strong, large-scale magnetic fields in early-type stars is still unknown and it has been suggested that stellar mergers may lead to the formation of magnetic stars \citep[e.g.,][]{ferrario2009,langer2012,w2014}. During mass transfer onto MS stars, the mass gainers rejuvenate because of mixing of fresh fuel into stellar cores. As a result of rejuvenation and increased masses, apparent stellar ages inferred from single-star models are younger than the real ages \citep[e.g.,][]{braun1995,bever1998,dray2007,fabian2014b,fabian2016}, potentially leading to $\tau$-distributions biased towards younger ages (depending on the exact star formation history). Contrarily, the $\tau$-distribution of a population of binary mass gainers, such as mergers, assuming a constant star formation history (i.e. for a given stellar mass, all ages and hence all fractional MS ages are equally probable) is biased towards older fractional MS ages. This can be understood as follows: extremely close binaries start mass transfer already close to the ZAMS and therefore add a uniform contribution to the overall $\tau$-distribution over the whole range of $\tau$. However most binaries need time before mass transfer starts, because the mass donors need time to evolve to fill their Roche lobes, resulting in uniform contributions that start at some fractional MS age and end at $\tau$\,=\,1. The starting point depends on the initial orbital separation and the exact rejuvenation. Summing the contributions of all possible mass gainers will therefore lead to a $\tau$-distribution that increases with fractional MS age. Binary mass transfer alone can therefore not explain the overall shape of the observed fractional $\tau$-distribution of magnetic stars.
\subsection{Magnetic suppression of core convection}\label{sec:core_convection}
It has been suggested that magnetic fields in the stellar core can stabilise convection, therefore reducing, or even stopping, convective core overshooting \citep{briquet2012}. \citet{petermann2015} suggested that the core of magnetic stars may even be smaller than that obtained without overshooting. Magnetic stars may therefore have smaller convective cores, hence shorter MS lifetimes (i.e. a narrower MS band in the HRD) than non-magnetic stars. If this is the case, the fractional MS ages of magnetic stars obtained using stellar evolution tracks that neglect convective core suppression would be systematically younger because of the wider MS band, possibly explaining the dearth of apparently old magnetic stars found in Sect.~\ref{sec:results}.

\begin{figure}[]
\includegraphics[width=85mm,clip]{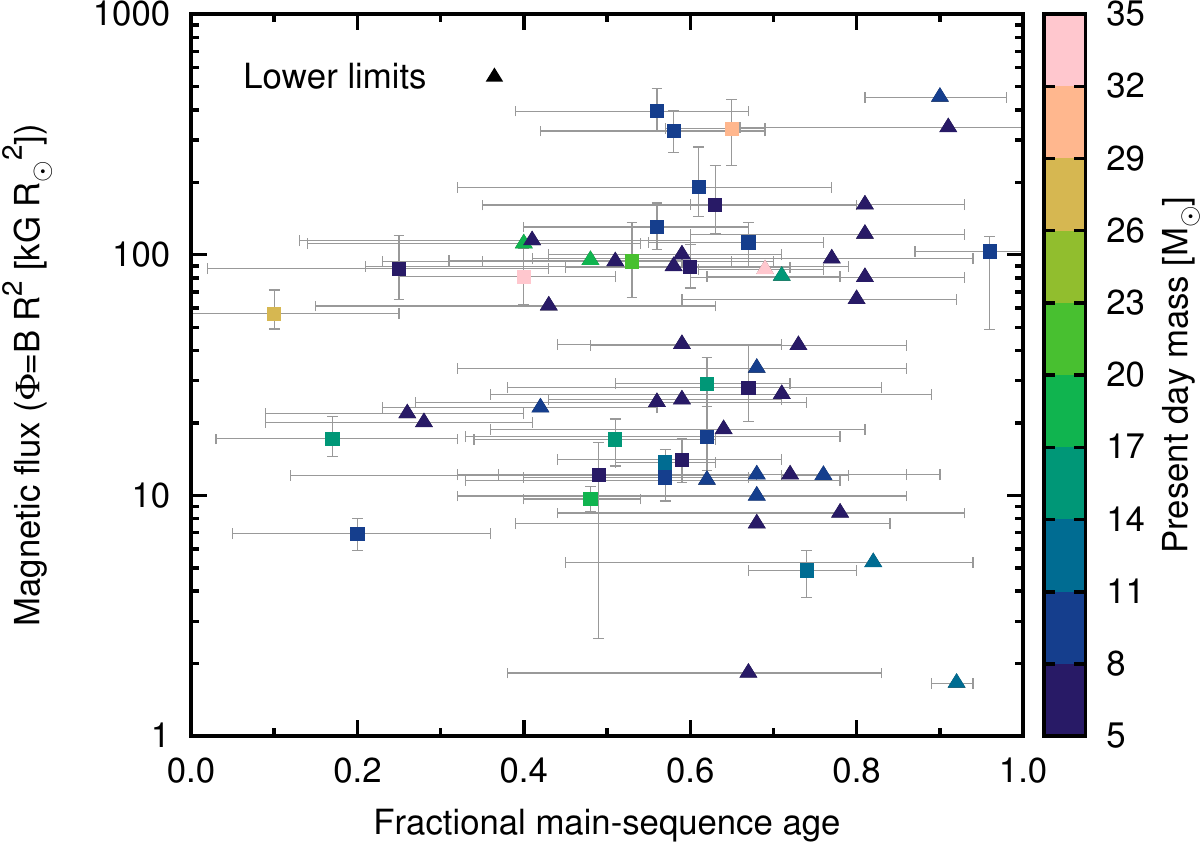}
\caption{Magnetic flux, $\phi$\,=\,B$_{\rm d}$R$^2$, as a function of fractional MS age for our sample of magnetic stars. The colour coding indicates the present-day masses of the stars. Triangles represent lower limits.}
\label{fig:magnetic-flux-vs-tau}
\end{figure}

If true, we might expect the suppression of core convection to increase with increasing magnetic field strength, although a saturation effect may exist. In that case, if the observed surface magnetic field is a good indicator of the field strength in the stellar core, then we expect a correlation between the derived fractional MS ages and the magnetic flux in the sense that apparently old magnetic stars must have weaker magnetic fields in the core and hence weaker magnetic fluxes on the surface in order to be detected as old with stellar models that neglect magnetic suppression of core convection. In Fig.~\ref{fig:magnetic-flux-vs-tau}, we show the magnetic flux of our sample of magnetic stars as a function of $\tau$. There is no clear evidence for a correlation between magnetic flux and $\tau$, suggesting that either magnetic fields do not hamper core convection to such a degree that we can observe clear differences in fractional MS ages or that the surface magnetic fluxes are not good indicators for the magnetic field strengths in stellar cores.
%
\subsection{Magnetic field decay}\label{sec:field_decay}
Magnetic fields may decay (i.e. surface magnetic field strength decreases below the detection limit) on a timescale that is of the order of the MS lifetime of massive stars. Such a decay would cause a decrease in the fraction of detected magnetic stars with increasing $\tau$, as observed in Fig.~\ref{fig:age-distribution}. We find that the dearth of evolved magnetic stars is possibly more pronounced at higher masses, suggesting that if field decay is mainly responsible for this dearth, it must occur on shorter time-scales in higher mass stars. Given that the dearth of evolved magnetic stars develops during the MS lifetime of stars, the field decay would have to occur on the nuclear timescale over a wide range of masses. The sharp decrease in the number of magnetic stars in the second half of the MS phase does not necessarily imply that magnetic field decay sets in at $\tau$\,$\approx$\,0.5. It is possible that field decay begins immediately after the generation and stabilisation of the magnetic field, but that the field remains on average strong enough to be clearly detectable up to half of the MS phase. In other words, it is only at $\tau$\,$\gtrsim$\,0.5 that magnetic fields start to become undetectable and it is only for the magnetic stars with the strongest initial fields that the magnetic field is still detectable at the TAMS.

Independent support for the conclusion that the magnetic fields of the known magnetic massive stars were possibly stronger in the past comes from the analysis of their spin-down ages (SD-ages; the time required to spin down a star from critical rotation to the current rotation rate). For each magnetic star in our sample, we derived the current SD-age following the formulation of \citet{udDoula2009} and \citet{petit2013}. We used the models by \citet{brott2011} to estimate the gyration constant, $k=(H/R)^2$ with $H$ being the radius of gyration and $R$ the stellar radius, following \citet{motz1952} for stars in the 3--40\,\M\ range at the ZAMS and TAMS, obtaining that it ranges between 0.002 (i.e. a 40\,\M\ star at the TAMS) and 0.07 (i.e. a 40\,\M\ star at the ZAMS). For all stars we adopted an average gyration constant of 0.04. The gyration constant and the mass-loss rate were assumed to be constant in time. For the calculation we adopted stellar parameters as described in Sect.~\ref{sec:sample}. For each star with \Teff\,$>$\,30000\,K, we derived the mass-loss rate using theoretical predictions by \citet{vink2000}, the same as adopted by \citet{brott2011}, while for cooler stars we adopted predictions by \citet{jiri2014}. We furthermore derived the stellar terminal wind velocity from the empirical relation given by \citet{castro2012} \citep[see also][]{kp2000}. For the stars for which just a lower limit on \bd\ and/or no rotation period was available, we derived an upper limit of the SD-age.
\begin{figure}[]
\vspace{-3.5cm}
\includegraphics[width=85mm,clip]{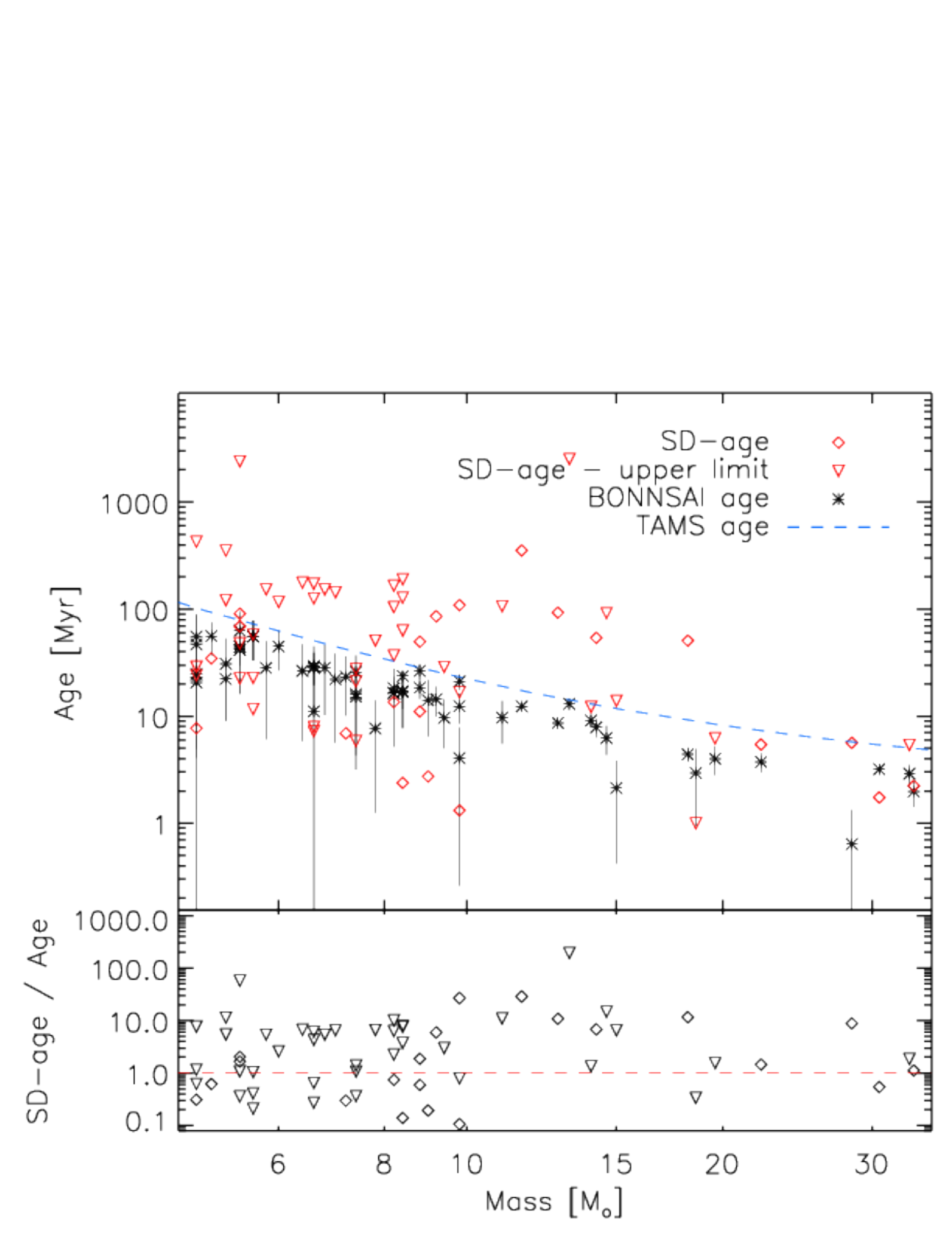}
\caption{Top: \bonnsai\ stellar ages for the sample of magnetic stars (black asterisks) and derived SD-ages (red rhombi and triangles) as a function of stellar mass. The red triangles indicate upper limits of the SD-age. The blue dashed line shows the TAMS age. Bottom: ratio between SD-ages and stellar ages as a function of mass. The red dashed line indicates equality of the two times.}
\label{fig:SD-time}
\end{figure}

Figure~\ref{fig:SD-time} compares the stellar ages derived with \bonnsai\ with the SD-ages for the magnetic stars in our sample. For half of the stars the SD-ages are larger than the stellar ages and in some cases even larger than the stellar lifetimes \citep[see also][]{morel2008,udDoula2008}. We tested this result against variations in the adopted terminal velocities, mass-loss rates, and moment of inertia constant. The use of different values of the terminal velocity, namely from \citet{kp2000}, does not affect the results. For most stars with masses lower than 10\,\M, the use of the mass-loss rates given by \citet{vink2000} leads to SD-ages smaller than the stellar ages, but the adopted mass-loss rates prescription of \citet{jiri2014} is more tuned to MS B-type stars, like the stars considered here. We also varied the moment of inertia constant and found that by using a value of 0.01, or lower, the SD-ages become smaller than the stellar ages for most stars.

With the adopted parameters, a past magnetic field strength that is at least about three times larger than the current one would be necessary to reconcile this discrepancy. This is based on the assumption that the magnetic field decays linearly with time, which may not be the case. 

\begin{figure}[]
\includegraphics[width=85mm,clip]{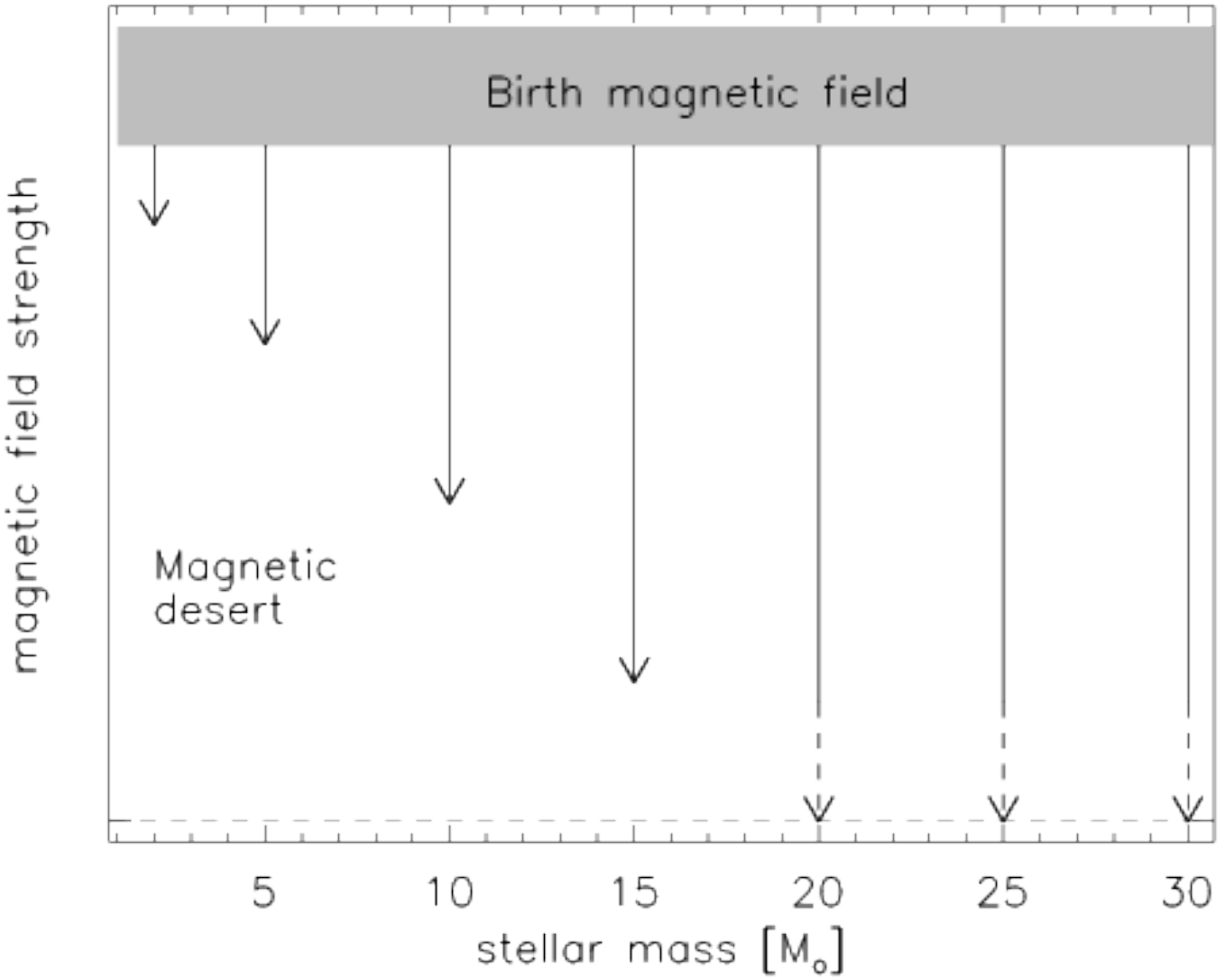}
\caption{Sketch of the decay in magnetic field strength as a function of stellar mass, according to our conjecture. The grey area at the top indicates the range of magnetic field values that stars might hold originally. Vertical lines, whose length is completely arbitrary, indicate time evolution, beginning at the time of magnetic field formation, and ending at the TAMS. The dashed sections of the lines indicate that it is not clear whether the magnetic field would completely disappear within the MS lifetime (see text). The dashed horizontal line indicates the level of no (or undetectable) magnetic field.}
\label{fig:cartoon}
\end{figure}

\citet{fossati2015} proposed that for massive stars there may be no magnetic desert in contrast to what is found for CP stars \citep[][found that there are no Ap stars with \bd\ values smaller than about 300\,G]{auriere2007}. In this context, a picture as outlined in Fig.~\ref{fig:cartoon}, where magnetic stars have birth fields in a discrete finite range and then decay on a timescale that decreases with increasing mass may well be compatible with the observed magnetic field strength distribution shown by \citet{fossati2015} and with the $\tau$-distribution of the fraction of magnetic stars (Fig.~\ref{fig:evolution-detection}). Within this picture, for the lowest mass stars, the decay is much slower than the MS lifetime, leading to the presence of a magnetic desert, while the magnetic field decay becomes comparable and then shorter than the MS lifetime with increasing mass. The different fractions of magnetic stars as a function of fractional MS age in the low- and high-mass stellar samples shown in Fig.~\ref{fig:age-distribution-2} would then be a natural consequence.

The strength of this scenario is that it leaves a unified picture for the magnetic fields in magnetic CP and massive stars, allowing for a magnetic desert in the former and a continuous distribution towards vanishing magnetic fields in massive stars. Figure~\ref{fig:cartoon} calls for a mechanism that allows fields to decay much faster in more massive stars. Perhaps this relates to the increased hydrodynamic diffusivity in massive stars as a consequence of the increasing importance of radiation pressure and the consequent weakening of mean molecular weight barriers \citep{potter2012,mitchell2015}.

A mass-dependent magnetic field decay would imply a lower detection rate for the most massive stars. \citet{wade2015a} reported that the detection rate in O-type stars is consistent with that of B-type stars, but this may be for example the result of an increased effect of binary mergers due to the higher rate of primordial massive O-type binaries \citep{sana2012}. Along this line, the picture in Fig.~\ref{fig:cartoon} might not be correct for all massive stars. For example, if magnetic fields for some stars were due to mergers, then such events occurring very late during core hydrogen burning might produce magnetic stars that have only a short time left to complete core hydrogen burning, such that their magnetic fields might not have time to fully decay \citep[see rejuvenation predictions of][]{fabian2016}.

Magnetic field decay implies that the number of stars that have host(ed) a magnetic field is larger than the measured incidence rate, as also suggested by \citet{fossati2015}. Massive stars present a very high component of slow rotators with \vsini\,$\lesssim$\,100\,\kms\ \citep[about 25\% in early B- and late O-type stars;][]{dufton2013,oscar2013,sergio2014a}, higher than the observed fraction of magnetic stars. Magnetic field decay, hence stars that are no longer detectable as magnetic, might help explain the difference between the observed incidence of magnetic fields and slow rotators in massive stars.
\section{Conclusions}\label{sec:conclusion}
From spectroscopically determined \Teff\ and \logg\ values, and with the aid of stellar evolution tracks and the \bonnsai\ tool, we derived a wide range of stellar parameters for a large sample of magnetic and non-magnetic massive stars. We showed that the fraction of magnetic massive stars remains constant up to a fractional MS age of $\approx$0.6 and decreases rapidly at older ages. By splitting the sample of stars according to their masses, we found that for the most massive stars ($>$14\,\M) the decrease in the rate of magnetic stars appears to be steeper and occurs at smaller fractional ages than for lower mass stars ($<$14\,\M).

We analysed the dipolar magnetic flux of our sample stars and found that flux conservation alone is not sufficient to explain the observed dearth of old magnetic massive stars. We also found that, in addition to flux conservation, neither binary rejuvenation in massive star mergers alone nor a suppression of core convection by the internal magnetic field would be able to explain the observed $\tau$-distributions. We therefore conclude that magnetic field decay is most likely the primary cause for the dearth of evolved magnetic stars. That field decay must occur is supported through our analysis of the spin-down ages of magnetic O- and B-type stars, showing that an original magnetic field strength of at least three times the current one is needed to reconcile the observed stellar ages and spin-down ages.

The differences between the $\tau$-distributions observed in the higher and lower mass magnetic samples suggest that the ratio of the field decay time to the stellar lifetime decreases with mass. On the one hand, this complies with the observed magnetic field strength distributions of intermediate-mass stars \citep{auriere2007,landstreet2007,landstreet2008}, where this ratio is higher than unity. This may explain the magnetic desert observed for these stars (cf. Fig.\,\ref{fig:cartoon}). On the other hand, this ratio becomes lower than unity in sufficiently massive stars such that - in contrast to intermediate mass stars - massive stars with weak fossil magnetic fields also exist \citep{fossati2015}. We speculate that the mass dependance of the magnetic field decay may be related to the less stable stratification in massive stars that is due to the increased fraction of radiation pressure \citep{potter2012,mitchell2015}. 

A magnetic field decay time shorter than the stellar lifetime does not automatically imply that the large-scale fields observed in massive MS stars play no role in their later evolution. If such fields were generated by stellar mergers, they may be produced in an advanced stage of core hydrogen burning in a considerable fraction of them. It is conceivable that in such cases, the field survives until core collapse and may turn the stellar remnant into a magnetar \citep{ferrario2008}. However, the magnetic fraction and flux distribution of magnetars can be expected to differ significantly from that observed in magnetic MS stars. Furthermore, a faster field decay in more massive stars might challenge the scenario that magnetars may form predominantly from very massive stars \citep{gaensler2005}.

These results and interpretations are based on the best of our current knowledge on massive star magnetism and evolution. It will be necessary to carry out more works to test our results, particularly after a larger number of homogeneous spectroscopic analyses of massive stars will become available and more magnetic massive stars are discovered, possibly also easing the constraint on the magnitude cut. This will increase the number of stars in both samples and decrease the effect of the biases, hence increase the sample representability and completeness.
\begin{acknowledgements}
LF acknowledges financial support from the Alexander von Humboldt Foundation. FRNS thanks the Hintze Family Charitable Foundation and Christ Church College for his Hintze and postdoctoral research fellowships. We thank John Landstreet for his useful comments on the manuscript. LF thanks Evelyne Alecian and Asif ud-Doula for fruitful discussions. GAW acknowledges support from an Natural Sciences and Engineering Research Council (NSERC) of Canada Discovery Grant. We thank the anonymous referee for the useful comments. Partially based on observations made with the Nordic Optical Telescope, operated by the Nordic Optical Telescope Scientific Association, and the Mercator Telescope, operated by the Flemish Community, both at the Observatorio del Roque de los Muchachos (La Palma, Spain) of the Instituto de Astrofisica de Canarias. SS-D acknowledges funding by the Spanish Ministry of Economy and Competitiveness (MINECO) under the grants AYA2010-21697-C05-01, AYA2012-39364-C02-01, and Severo Ochoa SEV-2011-0187, and by the Canary Islands Government under grant PID2010119. TM acknowledges financial support from Belspo for contract PRODEX GAIA-DPAC.
\end{acknowledgements}
%

%
%
%
%
%
%
\end{document}